\documentclass[final,1p]{elsarticle}

\usepackage{amssymb}
\usepackage{amsmath}

\usepackage{hyperref}
\usepackage{algorithm}
\usepackage{algpseudocode}
\usepackage{subfig}

\usepackage{lineno}

\journal{Digital Signal Processing}

\begin{document}

\begin{frontmatter}


\title{Direction of Arrival Correction through Speech Quality Feedback}

\author{Caleb Rascon} 
\ead{caleb@unam.mx} 

\affiliation{organization={Universidad Nacional Autonoma de Mexico},
            addressline={Circuito Escolar 3000}, 
            city={Mexico},
            postcode={03740}, 
            state={CDMX},
            country={Mexico}}

\begin{abstract}
	Real-time speech enhancement has began to rise in performance, and the Demucs Denoiser model has recently demonstrated strong performance in multiple-speech-source scenarios when accompanied by a location-based speech target selection strategy. However, it has shown to be sensitive to errors in the direction-of-arrival (DOA) estimation. In this work, a DOA correction scheme is proposed that uses the real-time estimated speech quality of its enhanced output as the observed variable in an Adam-based optimization feedback loop to find the correct DOA. In spite of the high variability of the speech quality estimation, the proposed system is able to correct in real-time an error of up to $15^o$ using only the speech quality as its guide. Several insights are provided for future versions of the proposed system to speed up convergence and further reduce the speech quality estimation variability.
\end{abstract}




\end{frontmatter}



\section{Introduction}

In real-life scenarios where a speech source of interest is present, and is of interest to process and analyze, various other audio sources typically coexist in the acoustic environment. This results in a mixture of the target speech source and these other sources, referred here to as `interferences', being captured in conjunction. Additionally, there are other effects that occur in real-life scenarios, such as noise and reverberation. All of these are aimed to be removed from the mixture, such that only the target speech source remains. This task, known as "speech enhancement," has shown significant advancements through deep learning methods \cite{das2021fundamentals}.

When conducted offline (using previously recorded audio), speech enhancement has benefited various applications such as security \cite{eskimez2018front} and music production \cite{porov2018music,lopatka2016improving}. Additionally, there is interest in performing speech enhancement in an online manner (using live audio capture), since it holds promise for a diverse range of applications, including real-time automatic speech recognition \cite{shi2024waveform}, sound source localization in robotics \cite{rascon2017localization}, hearing aids \cite{green2022speech}, mobile communications \cite{zhang2021sensing}, and teleconferencing \cite{rao2021conferencingspeech}.

It should be noted that carrying out a process in an online manner and in ``real-time'' are usually distinguishable. Online processing only requires that the execution is carried out within a time frame shorter than the capture time. While real-time processing, in addition to the latter constraint, also requires to meet specific latency or response time requirements that are application-dependent \cite{leinbaugh1980guaranteed,joseph1986finding}. For the sake of simplicity, this work focuses solely on meeting the execution-time-less-than-capture-time criterion, and the terms ``online processing'' and ``real-time processing'' are used inter-changeably.

To this effect, the online speech enhancement technique know as Demucs Denoiser \cite{defossez20_interspeech} has recently demonstrated superior performance compared to several other advanced methods when running in real-time \cite{rascon2023characterization}. It achieves excellent enhancement results, typically achieving an output signal-to-interference ratio exceeding 20 dB, even when using very short window segments (0.064 s). This positions it as a representative example of current advancements in online speech enhancement technologies.

However, similar to other speech enhancement techniques, the Demucs Denoiser model operates under a crucial assumption: that there is only one speech source in the mixture. This assumption may not pose a significant issue in applications like mobile telecommunication \cite{zhang2021sensing} and teleconferencing \cite{rao2021conferencingspeech}, which typically involve one user speaking at a time. However, in contexts such as service robotics \cite{rascon2017localization} and hearing aids \cite{green2022speech}, where multiple speech sources are expected, using speech enhancement techniques may not be suitable. In such cases, ``speech separation'' techniques \cite{wang2018supervised} could be more appropriate, as they aim to separate multiple speech sources into distinct audio streams. However, while speech enhancement techniques can be executed in real-time, speech separation techniques as of yet are not typically suitable for online applications \cite{rascon2023characterization}.

It has been demonstrated that when inputting a mix of speech sources into contemporary online speech enhancement techniques, some algorithms isolate the speech mixture from non-speech sources, whereas others, such as Demucs Denoiser, prioritize enhancing the ``loudest'' speech source \cite{rascon2023characterization}. To this effect, the work in \cite{rascon2023target}, which this work is based on, proposed two target selection strategies so as to ``nudge'' Demucs Denoiser towards the speech source of interest. The strategy that generally outperformed the other was the one that selects the target speech through its location (specifically, its direction of arrival). However, it was found to be very sensitive to localization errors.

In this work, a direction-of-arrival (DOA) correction mechanism is proposed to supplement this previous work, based on maximizing the speech quality measured in the audio output. A feedback loop is implemented, where the speech quality is used by an optimization mechanism as the observed variable, and a `corrected' DOA as its controlled variable. This corrected DOA is fed back to the location-based speech enhancement to obtain a new speech quality estimation, repeating the process to find the correct DOA.

This work has the following structure: Section \ref{sec:method} details the proposed system and all of its modules; Section \ref{sec:results} shows how the proposed system works in a real-time scenario, it also discusses its limitations and insights for future versions; finally, conclusions and future work are presented in Section \ref{sec:conclusions}.

\section{Proposed System}
\label{sec:method}

\begin{figure}[ht]
	\centering
	\includegraphics[width=0.75\linewidth]{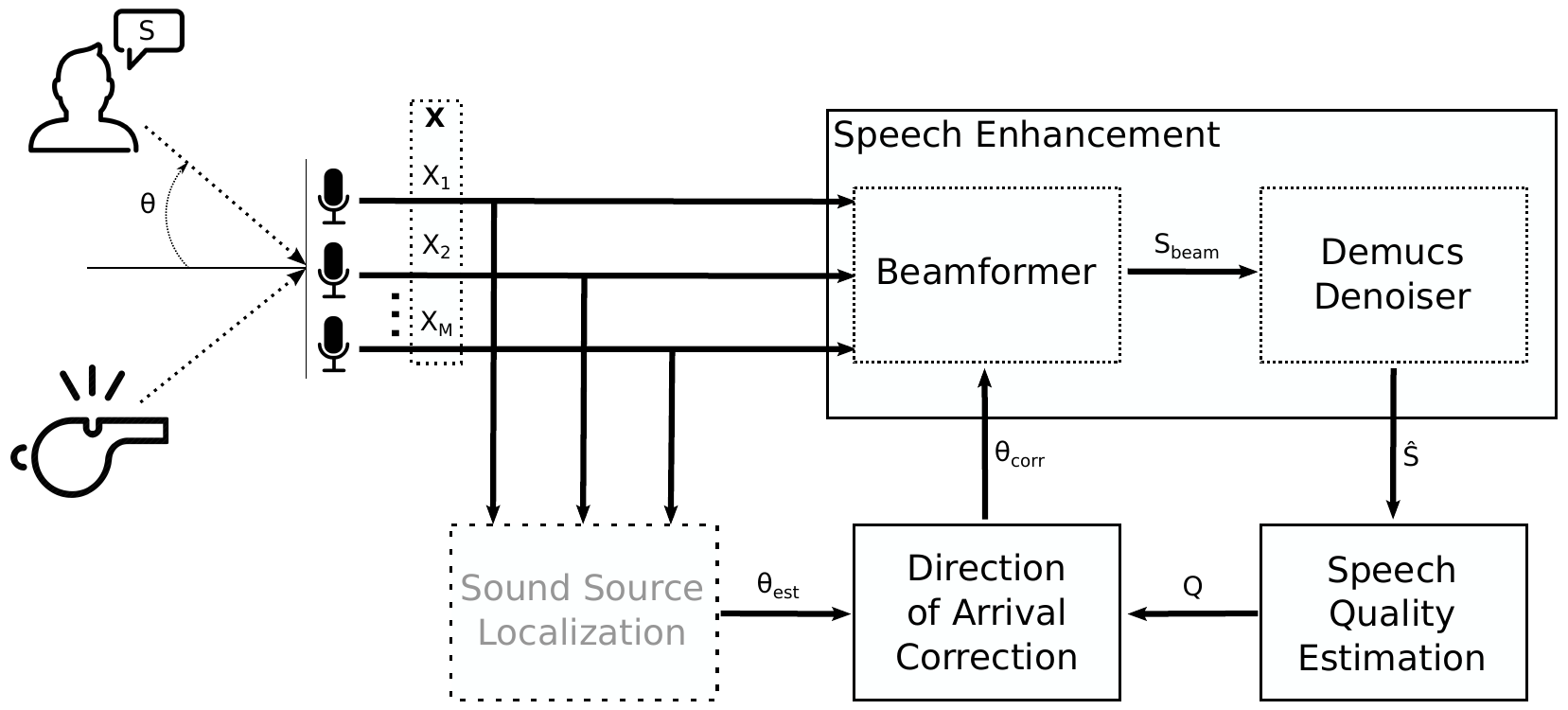}
	\caption{Overall proposed methodology.}\label{fig:proposed}
\end{figure}

The proposed system is presented in a general manner in Figure \ref{fig:proposed}. In summary, the microphone array input ($\boldsymbol{X} = [X_1, X_2, \cdot, X_M]$) is fed to both the speech enhancement module (detailed in Section \ref{subsec:se}) and a sound source localization technique. This technique is not part of the proposed system, since it is assumed that its localization estimation ($\theta_{est}$) may have errors that the proposed system aims to correct. The output of the speech enhancement ($\hat{S}$) is the output of the proposed system, which is fed back through the speech quality estimation module that provides a quality measurement ($Q$) of the system's output. This measurement is used by the direction-of-arrival correction module, in conjunction with the localization estimation ($\theta_{est}$), to provide a corrected direction-of-arrival ($\theta_{corr}$) to be used by the speech enhancement module, with the aim to provide a higher quality output.

In the following sections, these three modules are detailed. It is important to mention that, for reasons of reproducibility and ease of testing, the implementation of the proposed system is available at \url{https://github.com/balkce/doacorrection}, which includes the direction of arrival correction module, the speech quality estimation module, and the trained Demucs Denoiser sub-module. As for the beamformer sub-module, its implementation is available at \url{https://github.com/balkce/beamformer2}.

\subsection{Speech Enhancement}
\label{subsec:se}

The speech enhancement module is based on the location-based target selection strategy detailed in \cite{rascon2023target}, which for completeness sake is summarized here.

The Demucs Denoiser model \cite{defossez20_interspeech} has been shown to be effective at carrying out speech enhancement in an online manner, with relatively minimal computational power and a low response time \cite{rascon2023characterization}. However, in this later study \cite{rascon2023characterization} it was also shown that, as with any other current speech enhancement technique, the Demucs Denoiser model assumes that only one speech source is present in the input mixture. In the case of multiple speech source, it was shown that it tends to separate the ``loudest'' speech source in the input mixture. Thus, it requires a target selection strategy to appropriately select the speech source of interest (SOI). In \cite{rascon2023target}, two strategies were explored, and the one based on the location of the SOI provided good results.

The location-based strategy requires to know the location of the SOI in relation to the microphone array as a direction of arrival ($\theta$), which can be provided by a diverse set of sound source localization techniques \cite{rascon2017localization}. The target selection strategy uses a phase-based frequency-masking beamformer \cite{rascon2021corpus} to create a preliminary estimation of the SOI ($S_{beam}$). Although it may not be able to remove all the interferences (and even inserts some musical artifacts in its estimation), it does increase the energy of the SOI compared to the rest of the sound sources in the mixture. Such increase is enough so that the Demucs Denoiser model `picks it up' as the speech source to aim for and separate from the mixture.

However, in \cite{rascon2023target} it was also shown that the location-based target selection strategy is very sensitive to localization errors. A location error of 0.1 m resulted in a 5 dB drop in the average output signal-to-interference ratio (SIR), as well as a considerable increase in result variability. Several approaches were proposed as future work in \cite{rascon2023target} to circumvent this issue, such as:

\begin{itemize}
	\item Re-train the Demucs Denoiser model with data that is the result of artificially inserting location errors. It is to be noted that the author of \cite{rascon2023target} did attempt to do this, but the Demucs Denoiser model seemed to not be able to internally correct for such issues. However, no further tests were carried in this front (and are out of the scope of this work), which does mean that this approach may still merit exploring.
	\item Incorporate a robust localization method. Although it is also worth exploring, it does not fix any issues within the speech enhancement module; it just `pushes' the responsibility elsewhere.
	\item Incorporate a quality metric as feedback to correct the beamformer. This is the approach explored in this work.
\end{itemize}

To this effect, it is essential to have a quality metric to carry out this approach. This is detailed in the following section.

\subsection{Speech Quality Estimation}
\label{subsec:sqe}

The proposed system requires that the quality of the enhanced speech is measured in an online manner. This is a research problem onto itself, with two main challenges: 1) assessing the quality of the enhanced signal without a reference signal to compare it to; and 2) doing so in a window-by-window basis (since the whole input signal is unavailable). As for the first challenge, there have been several approaches that have attempted to solve it \cite{kumar2023torchaudio,fu2018quality,lo2019mosnet,dong2020pyramid,cauchi2019non,shen2023non}.

The approach employed in this work is the model known as \textit{Squim} \cite{kumar2023torchaudio}. The reasoning behind this selection is that: 1) it is one of the most recent, outperforming earlier approaches; 2) it is quite popular (and, thus, tested by the community) and already implemented as part of Torchaudio \cite{yang2022torchaudio}, the deep learning framework employed here; and 3) although it has not been reported as such, it is actually able to be run online (partially solving the second aforementioned challenge). To further this last point, its response times were measured as part of this work, using different capture window lengths ($t_w$), which are shown in Table \ref{tab:sqaresponse}.

\begin{table}[h]
	\centering
	\begin{tabular}{c c}
		Capture Time $t_w$ & Response Time \\
		(seconds) & [Min, Max] (seconds) \\
		\hline
		1.0 & [0.0234, 0.0242] \\
		2.0 & [0.0310, 0.0425] \\
		3.0 & [0.0538, 0.0704] \\
	\end{tabular}
	\caption{Squim response times.}\label{tab:sqaresponse}
\end{table}

Considering that these tests were carried out with a low-power GPU (Nvidia GTX 1050 Ti), these are very good response times. A time step ($t_h$) of 0.1 s is more than enough to measure the quality of the last 3.0 s of captured audio ($t_w$).

Additionally, Squim assumes that there is speech activity in its input, which may not always be the case. This can be solved by the use of voice activity detection as pre-processing step. To this effect, the Silero-VAD technique \cite{SileroVAD} was chosen because of its good performance, while providing low response times and requiring low computing power. To provide a smooth transition between quality estimations, the latest $t_w$ window is divided into smaller windows of length $t_{vad}$, and Silero-VAD is applied to each of them. If more than $3/4$ of the total amount of these $t_{vad}$ windows have active speech, the latest $t_w$ window is fed to Squim to obtain a quality estimation. This results in a series of continuous quality estimations through time, each calculated from the latest $t_w$ window of the input signal that has active speech.

\begin{figure}[ht]
	\centering
	\includegraphics[width=0.5\linewidth]{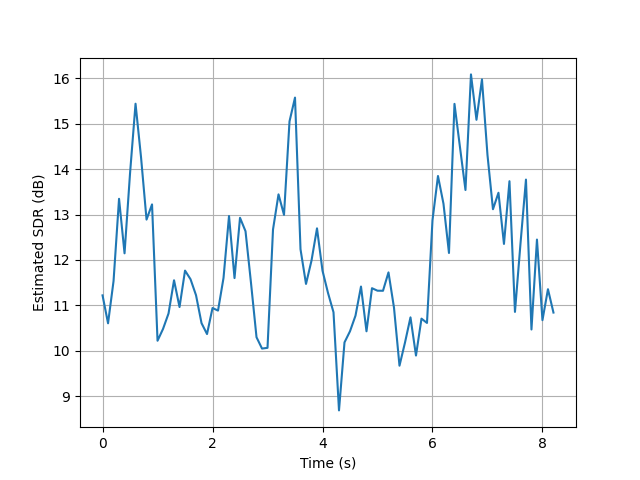}
	\caption{Quality estimated by Squim through time.}\label{fig:sqa_sdr}
\end{figure}

However, even when applying a VAD pre-processing step, Squim provides very `noisy' quality estimations through time, as shown in Figure \ref{fig:sqa_sdr} (with $t_h = 0.1$ and $t_w=3.0$).

\begin{figure}[ht]
	\centering
	\includegraphics[width=0.5\linewidth]{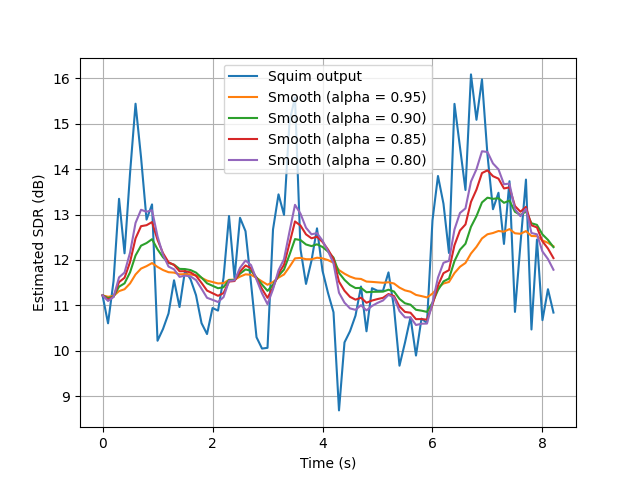}
	\caption{Squim smoothened through time.}\label{fig:sqa_smooth}
\end{figure}

As it will detailed later, the direction of arrival correction module works well with a `smooth' input/observed signal, which is not the case with the Squim output. To overcome this issue, exponential smoothing is applied, as presented in (\ref{eq:expsmooth}).

\begin{equation}
\label{eq:expsmooth}
Q_k \gets \alpha Q_k + (1-\alpha)Q_{k-1}
\end{equation}

where $Q_k$ is the quality estimation at the $k$ moment in time, and $\alpha$ is a smoothing factor in the range of $[0,1]$ with higher values providing smoother results but less responsiveness to underlying changes, and vice-versa. In Figure \ref{fig:sqa_smooth}, results are shown when applying different values of $\alpha$ to the quality estimations through time.

It can be seen that in all cases the final smoothened result is considerably less variable than the original Squim output. The $\alpha$ value of 0.9 is a good balance between providing a smooth output and still being somewhat responsive to underlying changes.

The complete speech quality estimation module is shown in Algorithm \ref{alg:squim}.

\begin{algorithm}[ht]
	\begin{algorithmic}[1]
		\Require $t_h$: step length
		\Require $t_w$: capture time
		\Require $t_{vad}$: VAD window length
		\Require $\alpha$: smoothing factor
		\State $n_{total} \gets \frac{t_w}{t_{vad}}$ \Comment{total number of $t_{vad}$ windows in $t_w$ window}
		\State Initialize: $k \gets 0$
		\Loop
		\State wait $t_h$ seconds
		\State $w_k \gets$ latest captured $t_w$ window
		\State $n_{act} \gets $ Silero-VAD$(w_k)$ \Comment{number of active $t_{vad}$ windows in $w_k$}
		\If{$n_{act} > \frac{3}{4}n_{total}$}
			\State $Q_k \gets $Squim$(w_k)$
			\State $Q_k \gets \alpha Q_k + (1-\alpha)Q_{k-1}$
			\State $Q \gets Q_k$
		\EndIf
		\State $Q \to $ direction of arrival correction module \Comment{output $Q$}
		\State $k \gets k + 1$
		\EndLoop
		\caption{Online speech quality estimation.}\label{alg:squim}
	\end{algorithmic}
\end{algorithm}

\subsection{Direction of Arrival Correction}
\label{subsec:doac}

Once the quality ($Q$) of a given past audio window has been provided, the direction of arrival correction module uses such information to correct the localization ($\theta_{est}$) estimated by a sound source localization technique. It does this by aiming to find the direction of arrival ($\theta_{corr}$) that maximizes $Q$.

The Adam optimization method \cite{kingma2014adam} is very popular in the deep learning community \cite{zhang2018improved}, typically used to find the weights of a given model that maximizes its performance. It has been mathematically proven to converge a non-convex objective function \cite{bock2019proof}, which is of great relevance to this work, given the estimated quality variability shown in Figure \ref{fig:sqa_smooth}. For this to be the case, however, it is required that the objective function be twice continuously differentiable. Unfortunately, this is not something that is ensured by purely smoothing the output of the speech quality estimation module (as described in the previous section). However, it does seem to provide an objective function that is `close' to satisfying such requirement.

Additionally, Adam is well fit to carry out its optimization process in an online manner, and because of its simplicity, a low response time can be assumed. It dynamically changes the updating factor of the controlled variable (which in this case is $\theta$) during its optimization process, considering the gradient of the optimized value (which in this case is $Q$). Adam then proceeds to use the first and second derivative of past gradients, by way of an exponentially decaying average. Both derivatives, respectively, correspond to the gradient's mean (or `momentum') and its uncentered variance. All of this in conjunction results in it avoiding `getting stuck' in local optima.

The Adam-based quality optimization process is presented in Algorithm \ref{alg:adam}.

\begin{algorithm}[ht]
	\begin{algorithmic}[1]
		\Require $t_h$: step length
		\Require $\beta_m, \beta_v$: forgetting factors for momentum ($m$) and variance ($v$)
		\Require $\eta$: learning rate
		\State Initialize: $\nabla m \gets 0$
		\State Initialize: $\nabla v \gets 0$
		\State Initialize: $\theta_p \gets 0$ \Comment{force an appropriate gradient at the start}
		\State Initialize: $\theta_c \gets \theta_{est}$ \Comment{$\theta_{est}$ from a sound source localization technique}
		\State Initialize: $Q_p \gets 0$
		\State Initialize: $Q_c \gets 0$
		\Loop
			\State $\theta_c \to \theta:$ to speech enhancement module \Comment{output $\theta$}
			\State wait $t_h$ seconds
			\State $Q_p \gets Q_c$
			\State $Q_c \gets Q: $ from speech quality estimation module
			\State $Q_c \gets 100 - Q_c$ \Comment{convert $Q_c$ to a minimizable value}
			\State $\nabla Q \gets \frac{Q_c-Q_p}{\theta_c-\theta_p + \epsilon}$ \Comment{calculate quality gradient}
			\State $\nabla m \gets \beta_m \nabla m + (1 - \beta_m)\nabla Q$ \Comment{update momentum and variance}
			\State $\nabla v \gets \beta_v \nabla v + (1 - \beta_v)\nabla Q^2$
			\State $\theta_p \gets \theta_c$
			\State $\theta_c \gets \theta_c - \eta \frac{\nabla m}{\sqrt{\nabla v} + \epsilon}$ \Comment{update $\theta$ with both gradients}
		\EndLoop
		\caption{Adam-based quality optimization.}\label{alg:adam}
	\end{algorithmic}
\end{algorithm}

As it can be seen, it requires the current and past quality estimations ($Q_c$ and $Q_p$, respectively), as well as the current and past direction of arrivals ($\theta_c$ and $\theta_p$, respectively). Three configurable parameters are meant to be set: the `forgetting' factors (as they are commonly known) for both the momentum and the variance ($\beta_m$ and $\beta_v$, respectively); as well as the learning rate ($\eta$) to update the direction of arrival.

It is important to mention that the Adam optimizer \cite{kingma2014adam} was originally intended as a minimization process. However, it is used here to maximize the speech quality. Thus, in the implementation shown in Algorithm \ref{alg:adam}, the result from the speech quality estimation module is subtracted from $100$ so that $Q_c$ bares a value that is aimed to be minimized.

It is also important to mention that the optimization process shown in Algorithm \ref{alg:adam} does not carry out the original Adam implementation, which usually includes bias correction to avoid `getting stuck' at the beginning of the optimization process. This is appropriate when the objective function is differentiable; this, in turn, implies that its current value can be robustly predicted given past values. However, as it was explained in the previous section, the output of the speech quality estimation module is not able to satisfy this assumption. Thus, in Algorithm \ref{alg:adam} no bias correction is carried out as part of the Adam-based optimization. 

The sound source localization estimation ($\theta_{est}$) is used at the starting point of the optimization process, established as the first value of $\theta_c$. Furthermore, $\theta_p$ is initialized at 0 so that the first calculated quality gradient has a `reasonable' value. If it is initialized with the same value as $\theta_p$, the first calculated quality gradient would be equivalent to $\frac{Q_c}{\epsilon}$, which results in an astronomical value. This, in turn, would make the mean and variance calculations have very similar values throughout the first iterations, effectively `halting' the optimization process in the mean time. It would only begin to update adequately until that first calculated gradient is `forgotten', which may take a considerable amount of iterations, given its enormous value.

It is also worth pointing out that the quality gradient ($\nabla Q$) is calculated using only the current and past data points. This means, that the instantaneous gradient is used. Additionally, $\theta_{est}$ is not used again during the optimization process. Although it would be worth exploring the impact of using more data points to calculate $\nabla Q$, as well as updated values of $\theta_{est}$ throughout the optimization process, it is left for future work. For the time being, the current proposal of the Adam-based optimization process seems to be enough to provide reasonable results, as it can be seen in the following section.

\section{Results}
\label{sec:results}

\subsection{Implementation and Evaluation}

A multi-channel recording from the AIRA corpus \cite{rascon2018acoustic} was used to test the proposed system. The recording bares two sound sources located at 1 m from the center of the microphone array, which has a triangular shape. There is a microphone at each vertex of the triangle; the inter-microphone distance is of 0.18 m. The source of interest is located at around $0^o$, and the interference is located at around $90^o$. The same recording was used throughout the tests so as to not have it be a source of variability that may impede comparability between results. 

A simple multi-channel reproduction program (referred here as \textit{ReadMicWavs}) was built using the JACK audio connection toolkit \cite{letz2005jackdmp} to feed the recording to the proposed system in real-time, so as to emulate a live microphone signal. Thus, all the results shown in the following sections are from tests that were carried out in an online manner. JACK was chosen given its well-standing performance to capture and reproduce multi-channel audio in real-time, with good inter-microphone synchronicity \cite{newmarch2017jack,letz2009s}.

To connect all of the previously described modules, ROS2 \cite{macenski2022robot} was used along with a custom message type called \textit{jackaudio} that holds: the window of audio data, its length and a timestamp. ROS2 was chosen since it has been widely used, mainly in the robotics and automation community \cite{reke2020self,erHos2019ros2}, for near real-time communication between modules, and has shown great potential to be used with low-power hardware \cite{maruyama2016exploring}.

As mentioned before, the beamformer used is the phase-based frequency masking technique \cite{rascon2021corpus}, implemented as part of the \textit{beamform2} ROS2 package (which can be accessed at \url{https://github.com/balkce/beamformer2}).

\textit{ReadMicWavs} uses the transport protocol in JACK to feed the recording (as if it were a real-time captured signal) to \textit{beamform2}, which acts as the beamformer sub-module shown in Figure \ref{fig:proposed}. This sub-module uses the direction of arrival (that was fed to it through the \textit{theta} ROS2 topic) to spatially filter the sound source of interest. Its results are published through the \textit{jackaudio} ROS2 topic (using the custom message type of the same name), which are fed to the \textit{demucs} ROS2 node which acts as the Demucs Denoiser sub-module in Figure \ref{fig:proposed}. The results from \textit{beamform2} are enhanced by \textit{demucs}, which are published through the \textit{jackaudio\_filtered} ROS2 topic (using the \textit{jackaudio} custom message type). These results are fed to the \textit{online\_sqa} ROS2 node, which acts as the speech quality estimation module in Figure \ref{fig:proposed}. Its quality estimations are published through the \textit{SDR} ROS2 topic and are fed to the \textit{doacorrect} ROS2 node, which acts as the direction of arrival correction module in Figure \ref{fig:proposed}. This module optimizes the speech quality by modifying the direction of arrival, which is published through the aforementioned \textit{theta} topic, closing the loop to the \textit{beamform2} node.

Because of the multi-modular nature of ROS2, the \textit{theta} topic can also be published by a node other than \textit{doacorrect}, which provides flexibility for future versions of the proposed system.

The values of the parameters (detailed previously) during testing are as follows:

\begin{itemize}
	\item Capture time ($t_w$): 3.0 s
	\item Time step ($t_h$): 0.1 s
	\item VAD window size ($t_{vad}$): 0.032 s
	\item Smoothing factor ($\alpha$): 0.9
	\item Momentum `forgetting' factor ($\beta_m$): 0.9
	\item Variance `forgetting' factor ($\beta_v$): 0.999
\end{itemize}

$t_w$ and $t_h$ were chosen considering the information shown in Table \ref{tab:sqaresponse}. $\beta_m$ and $\beta_v$ were chosen based on the recommendation in \cite{kingma2014adam}, that are typically used in other works as well. $\alpha$ was chosen considering what is discussed in Section \ref{subsec:sqe}.

To observe the overall repeatability of the proposed system's behavior, several `runs' were carried out with each configuration. Additionally, as it will be seen, several of these `runs' do not provide an appropriate behavior: the system gets `lost' and the corrected DOA is not near the correct DOA. To this effect, a `good' run is here defined as one that in its last third of its run-time, the average $\theta$ is less than $5^o$ from the correct DOA.

\subsection{Effect of Learning Rate ($\eta$)}

\begin{figure}[ht]
	\centering
	\subfloat[$\eta=0.01$.]{
		\includegraphics[width=0.45\linewidth,height=0.2\textheight]{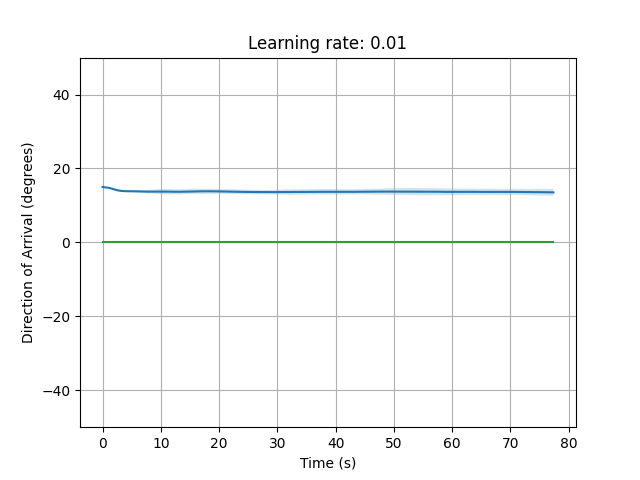}
		\label{fig:eta_a}
	}\qquad
	\subfloat[$\eta=0.1$.]{
		\includegraphics[width=0.45\linewidth,height=0.2\textheight]{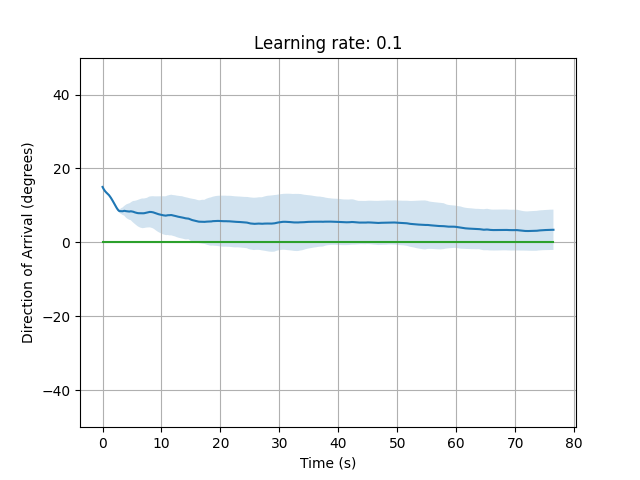}
		\label{fig:eta_b}
	}\qquad
	\subfloat[$\eta=0.2$.]{
		\includegraphics[width=0.45\linewidth,height=0.2\textheight]{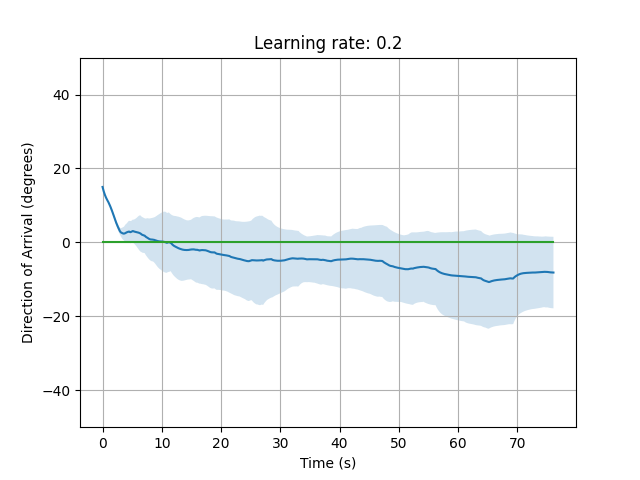}
		\label{fig:eta_c}
	}\qquad
	\subfloat[$\eta=0.3$.]{
		\includegraphics[width=0.45\linewidth,height=0.2\textheight]{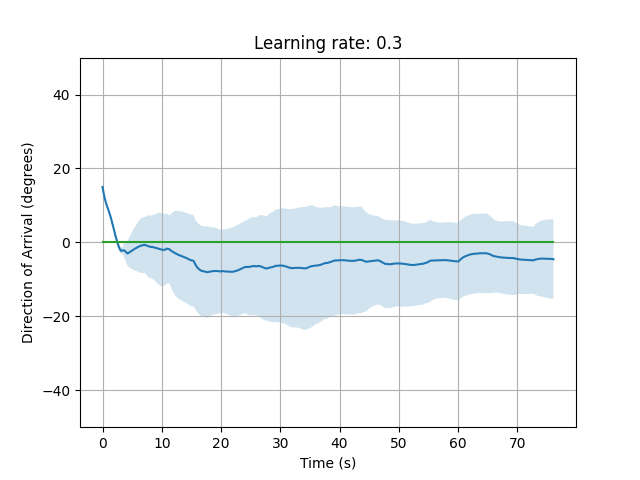}
		\label{fig:eta_d}
	}
	\caption{Effect of $\eta$. 5 runs with $\theta_{est}=15^o$.}
	\label{fig:eta}
\end{figure}

The learning rate parameter ($\eta$) is one of the cornerstones of the proposed system. To observe its effect, several runs were carried out using different values. The results are shown in Figure \ref{fig:eta}, where the dark blue line is the mean $\theta$ at each moment in time, and the space below and above such line represents the standard deviation (as a measure of variability).

From these results, it can be surmised that a low $\eta$ value (such as 0.01, shown in Figure \ref{fig:est_a}) gets `stuck' early on the optimization process, but it provides more consistent (less variable) results. On the other hand, a high $\eta$ value (such as 0.3, shown in Figure \ref{fig:est_d}) does not get `stuck' as easily, but provides much less consistent (more variable) results and tends to overshoot the target. A good balance between these two scenarios is a $\eta$ value of 0.1 (shown in Figure \ref{fig:eta_b}) which provides consistent results, while not getting `stuck' at the start and not overshooting the target.

\subsection{Effect of Bias Correction Removal}

\begin{figure}[ht]
	\centering
	\subfloat[With bias correction.]{
		\includegraphics[width=0.45\linewidth]{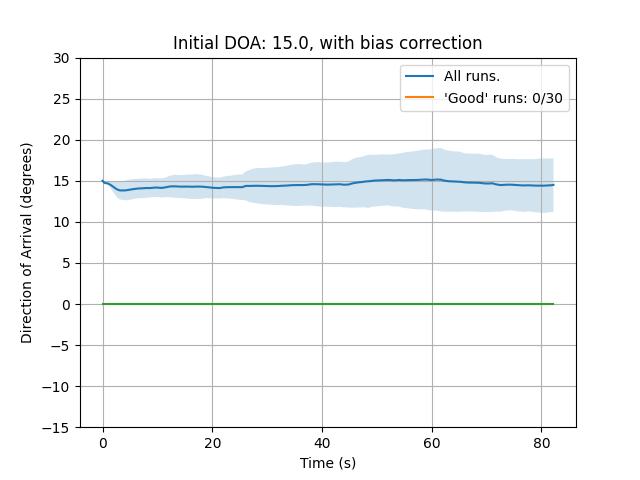}
		\label{fig:bc_a}
	}\qquad
	\subfloat[Without bias correction.]{
		\includegraphics[width=0.45\linewidth]{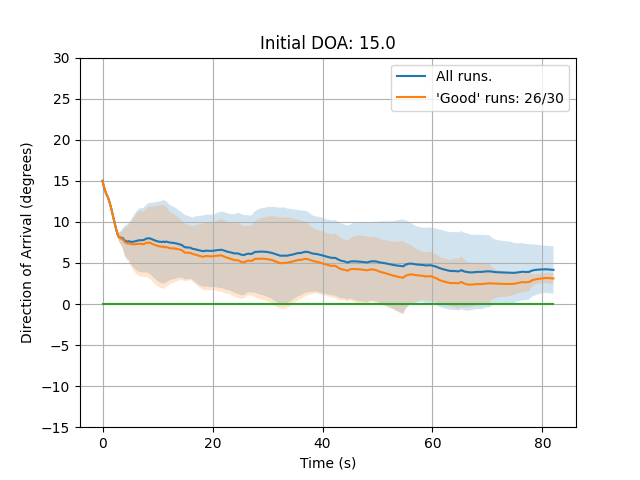}
		\label{fig:bc_b}
	}
	\caption{Effect of bias correction. 30 runs with $\theta_{est}=15^o$.}
	\label{fig:bc}
\end{figure}

To validate the removal of the bias correction in the Adam-based optimization shown in Algorithm \ref{alg:adam}, two tests were carried out as previously described, the results of which are shown in Figure \ref{fig:bc}. The difference between these two tests is that one was carried out using bias correction (Figure \ref{fig:bc_a}), and the other without (Figure \ref{fig:bc_b}).

As it can be seen, when applying bias correction, no `good' runs were observed and no tendency towards the correct DOA can be observed. This is opposed to when no bias correction is carried out, in which a considerably amount of `good' runs are observed and the overall tendency of the system is toward the correct DOA. This demonstrates that the overall response of the proposed system is improved when not applying bias correction in the Adam-based quality optimization.

\subsection{Effect of Estimated Direction of Arrival ($\theta_{est}$)}

\begin{figure}[ht]
	\centering
	\subfloat[$\theta_{est}=1^o$.]{
		\includegraphics[width=0.45\linewidth,height=0.2\textheight]{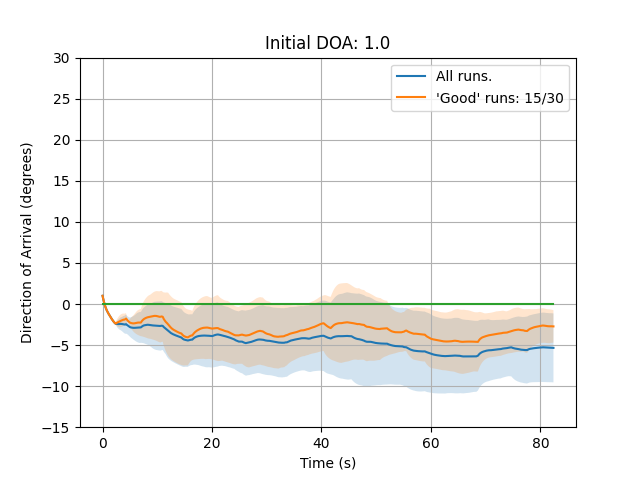}
		\label{fig:est_a}
	}\qquad
	\subfloat[$\theta_{est}=5^o$.]{
		\includegraphics[width=0.45\linewidth,height=0.2\textheight]{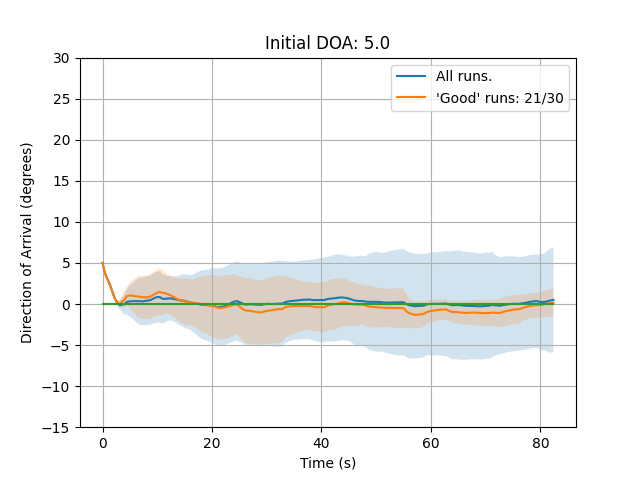}
		\label{fig:est_b}
	}\qquad
	\subfloat[$\theta_{est}=10^o$.]{
		\includegraphics[width=0.45\linewidth,height=0.2\textheight]{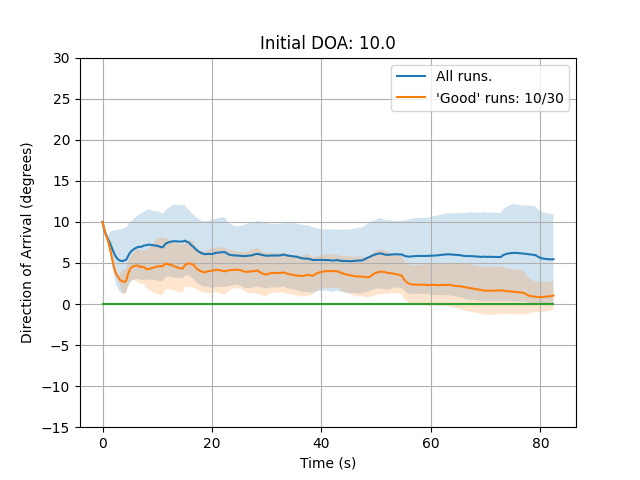}
		\label{fig:est_c}
	}\qquad
	\subfloat[$\theta_{est}=15^o$.]{
		\includegraphics[width=0.45\linewidth,height=0.2\textheight]{doaopt_various-15.0}
		\label{fig:est_d}
	}\qquad
	\subfloat[$\theta_{est}=20^o$.]{
		\includegraphics[width=0.45\linewidth,height=0.2\textheight]{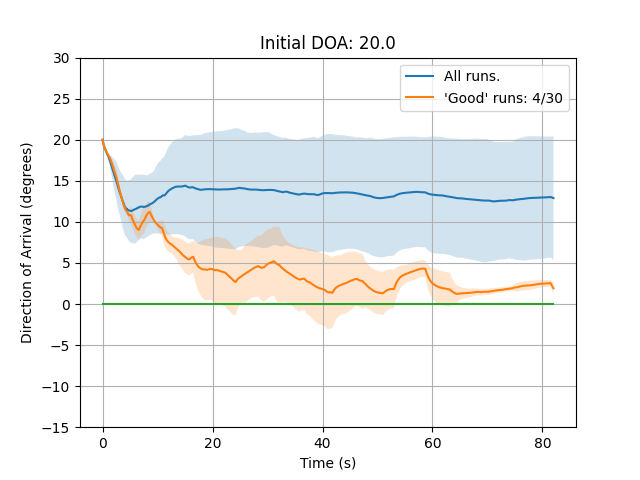}
		\label{fig:est_e}
	}\qquad
	\subfloat[$\theta_{est}=25^o$.]{
		\includegraphics[width=0.45\linewidth,height=0.2\textheight]{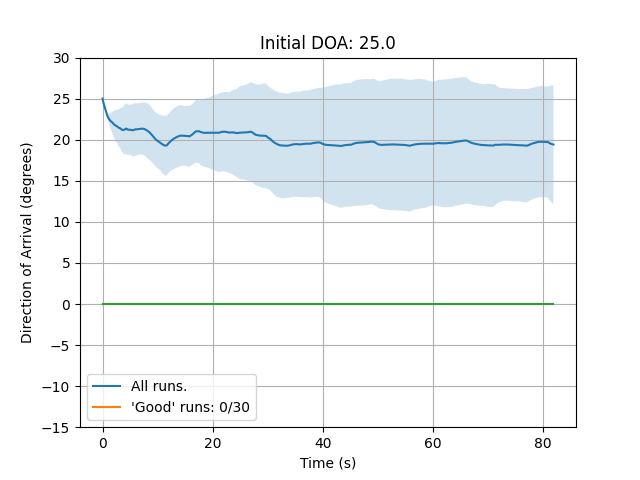}
		\label{fig:est_f}
	}
	\caption{Effect of $\theta_{est}$. 30 runs with $\eta=0.1$.}
	\label{fig:est}
\end{figure}

To observe the effect of the estimated direction of arrival, several runs were carried out varying the estimated direction of arrival ($\theta_{est}$), and the results are shown in Figure \ref{fig:est}. This variation simulates scenarios in which a sound source localization technique provides an estimated direction of arrival with varying degrees of error.

\newpage

As it can be seen, the proposed system is able to correct the DOA when $\theta_{est}$ is less than $20^o$ away from the correct DOA. From $20^o$ onward, the amount of `good' runs is reduced considerably, with $\theta_{est}=25^o$ being the limit with which it is not able to `recover'. Additionally, it can also be seen that as the error in $\theta_{est}$ increases, the more time it takes the proposed system to reach a value near the correct DOA.

It is worth pointing out that when $\theta_{est}=1^o$, the proposed system actually inserts a DOA error instead of correcting it, even though its starting DOA is close to the correct DOA. It will be discussed further in Section \ref{subsec:discuss}, but this is due to the fact that the system is not able to converge in a single value (because of the input variability and the nature of the optimization approach). Future efforts are to be made so that the proposed system behavior is more stable.

Another noteworthy case is when $\theta_{est}=15^o$, which is an outlier of the following tendency: as $\theta_{est}$ gets farther from the correct DOA, the lower amount of `good' runs. This is due to the fact that the proposed system's behavior starts with a noticeable downward `push' (which is the result of the $\theta_p \gets 0$ step in Algorithm \ref{alg:squim}), and that its size is dependent on the value of the first quality estimation. In fact, a tendency of the size of this downward can be observed in Figure \ref{fig:est}, where the closer $\theta_{est}$ is to the correct DOA, the lower this downward `push'. This turned out to specifically benefit the case when $\theta_{est}=15^o$, since that initial downward `push' makes the proposed system land close enough to the correct DOA (probably just inside the valley of the global optimum in the search space) such that the subsequent DOA correction requires only to refine its result. Its important to mention that the value of $\eta$ was chosen while $\theta_{est}$ was set at $15^o$, so this is evidence of a type of over-fitting of the system's behavior. As it will be discussed in Section \ref{subsec:discuss}, it is then of great interest to explore an automatic parameter calibration scheme that involves the selection of all the parameters in conjunction so as to obtain a more generalizable optimization behavior. Having said all of this, the combination of $\alpha=0.9$ and $\eta=0.1$ seems to be providing an acceptable optimization performance given that $\theta_{est}<20^o$.

\subsection{Other Direction of Arrivals and More Intereferences}

\begin{figure}[ht]
	\centering
	\includegraphics[width=0.5\linewidth]{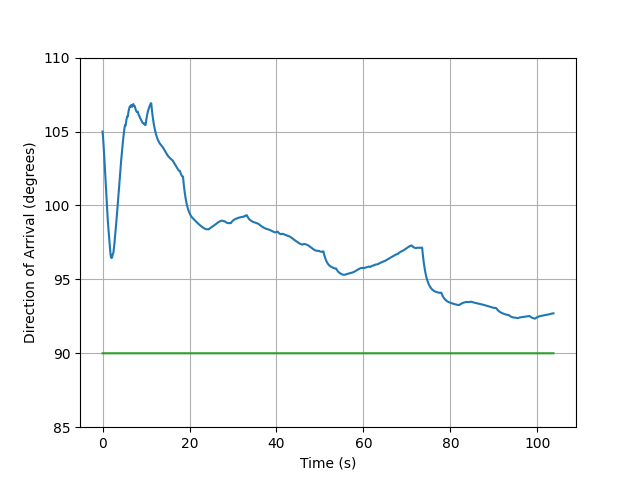}
	\caption{Result when $\theta_{est}=105^o$, near the other source located at $90^o$.}\label{fig:est90}
\end{figure}

To further inspect the generalizability of the proposed system, another test was carried out when $\theta_{est}=105^o$, which is closer to the other source located at $90^o$. The result is shown in Figure \ref{fig:est90}.

As it can be seen, the proposed system got close ($\pm 5^o$) to the correct DOA.

\begin{figure}[ht]
	\centering
	\subfloat[$\theta_{est}=105^o$.]{
		\includegraphics[width=0.45\linewidth,height=0.2\textheight]{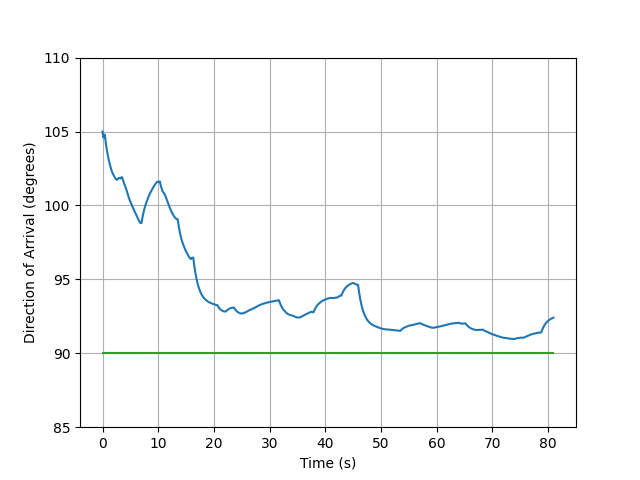}
		\label{fig:noisy_a}
	}\qquad
	\subfloat[$\theta_{est}=195^o$.]{
		\includegraphics[width=0.45\linewidth,height=0.2\textheight]{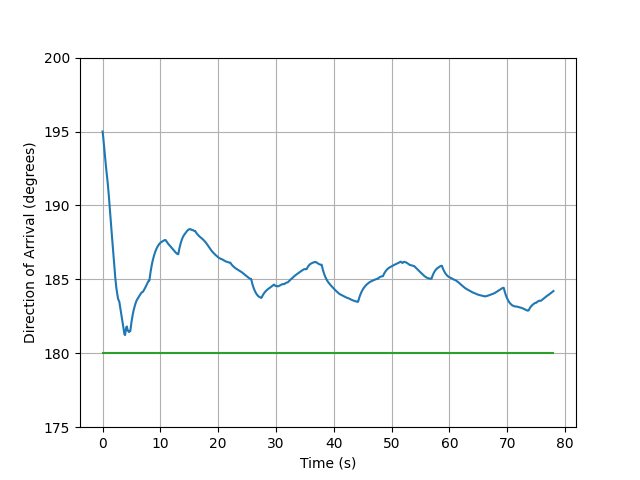}
		\label{fig:noisy_b}
	}\qquad
	\subfloat[$\theta_{est}=-15^o$.]{
		\includegraphics[width=0.45\linewidth,height=0.2\textheight]{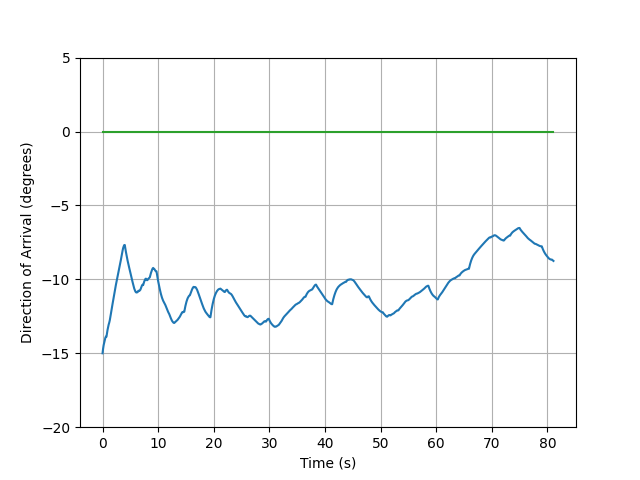}
		\label{fig:noisy_c}
	}
	\caption{Results of running with more interferences.}
	\label{fig:noisy}
\end{figure}

Additionally, it is also of interest to explore the proposed system's performance with more interferences present. In Figure \ref{fig:noisy}, the results are shown of a 3-source scenario, one located at around $90^o$, another at around $180^o$, and the third at around $0^o$. The three runs, respectively, had $\theta_{est}=105^o$, $\theta_{est}=195^o$, and $\theta_{est}=-15^o$ (a $15^o$ initial error). As it can be seen, the two first runs got close ($\pm 5^o$) to the correct DOA. The second run didn't got as close, but its definitely trending towards the correct DOA.

\newpage

\subsection{Discussion}
\label{subsec:discuss}

As far as the author knows, this is the first successful attempt to carry out real-time direction-of-arrival correction by only using as feedback the estimated speech quality. Overall, the proposed system is performing this task well. However, there are some issues that need to be considered for later versions of the proposed system.

First off, the system can be considered slow. In general, a `good' run takes up to 40 seconds (as shown in Figure \ref{fig:est_d}) to reach a value near the correct DOA. Unfortunately, accelerating the optimization process requires increasing the learning rate, which results in the system getting `lost' more frequently. Also, the proposed system seems to be sensitive to the initial circumstances in which the DOA correction was carried out (which is the reason why several runs were needed to be carried out for each test). It is the belief of the author that this is due to the high variability of the quality estimations. The causes of this may be many-fold. For one, the quality estimations are initially provided by Squim which are highly variable. Additionally, although ROS2 is presumed to be able to run in real-time, the manner in which it interacts with a true real-time server, such as JACK, may result in an overrun: where the execution time of a given input data window ends up being higher than its capture time. The rate of this occurrence appears to be agnostic to the response time of the system (which is quite low, as will be discussed later); a possible reason is that the internal communication mechanisms of ROS2 are not built to be truly run in real-time. In any case, when an overrun occurs, a zero-valued output data window is returned, resulting in small ``breaks'' (or silences) in the enhanced signal. The quality estimation is then affected, resulting in more variability. All of this in conjunction makes it difficult for the optimization process to converge in a single value. Instead, a slight but continuous variation of the corrected DOA is observed, the average of which is usually close ($\pm 5^o$) to the correct DOA. However, in cases when the initial DOA estimated by the sound source localization technique is already close to the correct DOA (as in the case of Figure \ref{fig:est_a}), the optimization process will introduce errors in an already close-to-correct estimated DOA.

In addition, it was observed that when the source was not well enhanced in its ideal state (when accurately located from the start), the quality estimation objective function provided little to no quality improvement compared to other values of $\theta_{est}$. This resulted in a sparse search space that is difficult to optimize. Fortunately, in these cases, the behavior of the proposed system still `tended' towards the correct DOA (exemplified in Figure \ref{fig:noisy_c}), which implies the need for more time to converge.

Thus, the optimization process would benefit from using other types of approaches, such as those applied in control engineering \cite{wang2009fast,shtessel2014sliding}, to handle the input variability while being fast to converge in a single steady value. Also, although substituting ROS2 as the inter-module communication framework is outside the scope of this work, it would be of interest to explore ways that make the real-time interaction between ROS2 and JACK be more seamless.

Next, it can be assumed that if and when the proposed system converges on the correct DOA, the speech quality is maximized. Thus, it can be argued that the proposed system is carrying out, simultaneously, both DOA correction and speech quality maximization. However, the latter was not discussed here because the main focus of this first version of the proposed system is DOA correction. Also, it is suspected that the speech quality improvement is minimal. To be fair, this assessment was obtained informally during the implementation of the proposed system, where the resulting speech was subjectively evaluated. Thus, it is left for future work to formally assess the speech quality improvement, as well as its impact in other possible subsequent steps in the audio processing data flow, such as sound source classification or speech recognition.

Furthermore, the only parameters whose impact on the system's behavior was characterized in this work was of $\eta$ and of $\alpha$. The former resulted in a type of over-fitting to the specific case of $\theta_{est}=15^o$ (as shown in Figure \ref{fig:est}), while the latter was admittedly very superficial and subjectively chosen. First off, the impact of the values of other parameters (such as $\beta_m$, and $\beta_v$) are also worth exploring. It would be of interest to observe how their values, in conjunction, impact the system's optimization behavior. Additionally, it would be of interest to create an automatic parameter calibration scheme that could run alongside the proposed system and provide a more dynamic and off-the-shelf application.

Moreover, it was observed how the bias correction in the Adam-based optimization was making the proposed system get `stuck' prematurely. This is counter-intuitive, since the main objective of the bias correction is to avoid Adam getting stuck at the beginning of the optimization process. However, given that the objective function is not differentiable, it was expected that the original Adam implementation would need modification to work well. Interestingly, since no bias correction is being carried out, this also means that the global optimum can be considered as dynamic, opposed to the assumption of global optimum staticity in the original Adam implementation. This opens the possibility to track mobile sound sources in later versions of the proposed system.

Additionally, it was observed that the Adam-based optimization was better served when the speech quality estimation module had run for a short while ($\sim 10$ s) before running the DOA correction module. It appears that the combination of the Squim model and the exponential smoothing requires some initial data to provide a `moderately stable' objective function signal. This, in turn, creates an initial state that avoids the Adam-based optimization getting `lost' at the start.

Finally, since the proposed system is aimed to be run in real-time, it is of interest to know its response time. Since all the modules shown in Figure \ref{fig:proposed} run in parallel, the overall response time is based solely on the `slowest' module. To this effect, the beamfomer sub-module runs as a client of JACK, which has a maximum latency of 0.021 s (as it was configured); during testing, this sub-module showed a response time close to half that amount, with a minimal amount of overruns occurring ($< 5$ in an hour long evaluation). The direction-of-arrival correction module runs a simple Adam-based optimization process, which bares a response time lower than 0.001 s. The speech enhancement sub-module runs the Demucs model that has a measured response time between 0.006 and 0.009 s to enhance a window of 0.063 s. And the speech quality estimation module runs the Squim model which, as shown in Table \ref{tab:sqaresponse}, has a measured response time between 0.0538 and 0.0704 s to measure the quality of a 3-second window. Since this last module is the slowest, the system's response time is between 0.0538 and 0.0704 s.

\section{Conclusions}
\label{sec:conclusions}

Speech enhancement carried in an online manner is of great interest to various areas of application. However, doing so using deep-learning-based models has shown to be challenging, given their low response time. To this effect, recent efforts have been made to not only making this type of models run in real-time while still providing comparable performance, as is the case of the Demucs Denoiser model, but also making them be able to focus on a given speech source of interest in a multi-speech environment. One of these efforts uses the location (mainly, the direction of arrival) of the source of interest to make Demucs Denoiser ``focus'' on it. However, this effort has been proven to be sensitive to location errors.

In this work, a direction-of-arrival correction system is proposed that is based on maximizing the speech quality of the speech enhancer. Through a feedback loop, the quality estimation is carried out via the Squim model, whose output is post-processed using exponential smoothing, to provide a close-to-differentiable objective function. An Adam-based optimization scheme is then applied to find the direction of arrival that maximizes such function.

It was shown that the proposed system works well, in that it is able to correct the direction of arrival towards the correct location. However, it was found that the system is sensitive to the learning rate of the Adam-based optimization, of which a recommended value was provided. Making the process less sensitive towards such value, as well as reducing the variability of the speech quality estimations, is of great interest and will be carried out as future work.

\section*{Acknowledgements}
The author would like to acknowledge the support of PAPIIT-UNAM through the grant IN100624.

\end{document}